\documentstyle[pre,aps,preprint,eqsecnum]{revtex}
\begin{document}
\draft
\title{
  Behavior of fluidized beds
  similar to equilibrium states}
\author{Kengo Ichiki\thanks{e-mail: ichiki@phys.h.kyoto-u.ac.jp}
  \& Hisao Hayakawa\thanks{e-mail: hisao@phys.h.kyoto-u.ac.jp}}
\address{Graduate School of Human and Environmental Studies,
  Kyoto University, Kyoto 606-01, Japan}
\date{\today}
\maketitle
\begin{abstract}
Systematic simulations are carried out
based on the model of fluidized beds proposed by the present authors
[K.Ichiki and H.Hayakawa, Phys. Rev. E {\bf 52}, 658 (1995)].
From our simulation, we confirm that
fluidization is a continuous transition.
We also confirm the existence of two types of fluidized phases, 
the channeling phase and the bubbling phase.
We find the close relations between
the averaged behaviors in fluidized beds
and quasi equilibrium states in dense liquids.
In fluidized beds,
(i) the flow rate plays the role of the effective temperature, and
(ii) the existence of a kind of the fluctuation-dissipation relation
is suggested.
\end{abstract}


\section{Introduction}
Recently granular materials have been studied extensively from both
experimental and theoretical point of views
in the context of the nonequilibrium statistical physics
\cite{jaeger1996b,mehta,hayakawa1995r}.
Since the granular materials are dissipative,
energy injections are necessary to preserve steady states.
Many of recent studies for granular materials
are focused on the behavior of systems excited by mechanical activations
such as vibration or rotation of vessels.
On the other hand,
the researches on fluidized beds,
where systems are excited by the fluid flow,
are not relatively advanced in spite of their variety of dynamical behaviors
\cite{davidson,gidaspow}.

Fluidized beds are widely used in chemical industries
since early 19's,
and they have been studied from technological point of views.
Fluidized beds consist of
granular particles confined in a tall chamber
with distributor for the fluid flow at the bottom.
In experiments, energy injection to the system
is controlled by the flow rate of fluid.
At low flow rate, the system is in the fixed phase
where particles rest on the bottom.
When the flow rate exceeds the critical value, particles start moving.
This state is called the fluidized phase,
which contains sub phases,
for instance, the homogeneous phase, the bubbling phase,
the channeling phase, etc.

There are many models to describe fluidized beds,
which can be classified in two categories,
two-fluid models and particle-dynamics models.
In the two-fluid models,
particles are treated as a fluid
\cite{drew1983,batchelor1988,sasa1992,goz1992,komatsu1993}.
These models have benefit on analytical treatments and
to generalize their discussion to other systems\cite{komatsu1995}.
However, their bases, such as
constitution equations for the particle-phase pressure and
the stress tensor, have not been established.
On the other hand,
the particle-dynamics models describe the direct motion of particles.
There are various models, which are
kinetic theories\cite{gidaspow},
the discrete element method\cite{tsuji1992,tsuji1993,tanaka1993}, etc.
However, these models cannot be the basis
for the two-fluid models.
The main problem of them is that
hydrodynamic interactions among particles
are over simplified.
For instance,
the boundary condition between particles and fluid
is not satisfied in the scale of particles,
and the fluid equation is calculated under inviscid limit.

Recently
the present authors have proposed a numerical model
of fluidized beds,
where hydrodynamic interaction among particles
is calculated with reliable accuracy\cite{ichiki1995,ichiki1996-phd}.
In this paper,
we will present the results of our systematic simulations
and behaviors on statistical quantities
obtained from the simulations.

The contents of this paper are as follows.
In Sec. II, we review the method of our simulation.
We show the results in Sec. III,
where we observe the transition of fluidization
and the existence of two fluidized phases.
We also discuss statistical quantities,
which are analogous to equilibrium correspondences.
In Sec. IV we give an interpretation of the averaged quantities
by the hole theory for simple liquids\cite{frenkel,hansen}.
In Sec. V, we conclude our results.
In Appendices, we summarize the method of our simulation
and discuss theoretical difficulties in the modeling of fluidized beds.

\section{Simulation method}
In this section, we briefly explain our model
and how to simulate the dynamics of granular particles in fluid flows.
The detail explanation of our model
can be seen in Ref.\cite{ichiki1995,ichiki1996-phd} and Appendices.
For simplicity, we only consider the cases of
monodispersed spherical particles.
We assume the following equation of motion for the particles
\begin{equation}
  \label{eq:eofm-efct}
  St
  {d{\bf U}\over dt}
  =
  -{\bf U}+{\bf V}+{\bf F}_c,
\end{equation}
where $St$ is the effective Stokes number,
${\bf F}_c$ represents hard-core collision among particles,
${\bf U}$ is the velocity of particles.
${\bf V}$ is the terminal velocity determined by
\begin{equation}
  \label{eq:term-vel}
  {\bf V}-{\bf u}^\infty
  =
  -\tensor{R}^{-1}\cdot
  {\bf E}_z,
\end{equation}
where ${\bf u}^\infty$ is the flow rate of induced fluid
which is equal to the superficial velocity conventionally used
for the fluidized beds.
$\tensor{R}$ is the resistance matrix representing
the hydrodynamic interaction among particles calculated by the method
of the Stokesian dynamics\cite{brady1988a},
where periodic boundary condition is adopted as the effect of chamber
(see Appendix \ref{sec:hydrodynamic}).
Hard-core collisions are assumed to be elastic
and calculated by the momentum exchange for contacting particles
in simulations (see Appendix \ref{sec:eofm}).
The bold-face letters without superscripts
represents vectors in $3N$-dimension,
where $N$ is the number of particles in the unit cell
of periodic boundary condition.
For example, the velocity ${\bf U}$ has the following components
\begin{equation}
  \label{eq:n-vector}
  {\bf U}
  =
  \left[
  \begin{array}{c}
    {\bf U}^{(1)}\\
    {\bf U}^{(2)}\\
    \vdots\\
    {\bf U}^{(N)}
  \end{array}
  \right],
\end{equation}
where the bold-face letters with superscripts
represent vectors in $3$-dimension.
In this paper, we use dimensionless quantities
with the aid of the particle radius $a$ and
the sedimentation velocity of a single particle in a viscous fluid
$U_0 = m\tilde{g}/6\pi\mu a$,
where
$\mu$ is the viscosity of the fluid,
$m$ is the mass of the particles and
$\tilde{g}=g(\rho_p-\rho_f)/\rho_p$
with
the gravitational acceleration $g$,
and the densities of the particle $\rho_p$ and the fluid $\rho_f$.
Equation (\ref{eq:eofm-efct}) represents the relaxation process
of ${\bf U}$ to ${\bf V}$ with the time-scale $St$.

In Eq.(\ref{eq:eofm-efct}),
there are two control parameters,
the effective Stokes number $St$ and the flow rate $u^\infty$.
For the parameters related to the system size,
we choose the number of mobile particles $N_{\text{M}}=256$,
that of fixed particles $N_{\text{F}}=10$
and the size of the unit cell in periodic boundary condition
$(L_x,L_y,L_z)=(34,2,100)$.
In this situation,
particles are confined in the vertical plane,
while hydrodynamic interactions are considered in 3-dimensional space.
We adopt the fixed phase as initial conditions of our simulations,
which is constructed from simulations with $u^\infty=0$.
The choice of the system size and these artifact situations
come from the limitation of computer resources.
We have checked that
statistical quantities seems to be insensitive to the choice of $L_z$
within the range of $50\leq L_z\leq 100$ and
the choice of the initial conditions is not relevant
from the comparison of results with other initial conditions.
We have also confirmed that
qualitatively similar behaviors to this situation
are observed in 3-dimensional simulations
and in the case of $N_{\text{M}}=133$.

\section{Simulation results}
\subsection{General behavior}
In this section, we present the results of our simulations in details.
We perform simulations at the points
in the parameter space (Fig. \ref{fig:parameter-space})
within the range of $0.05\leq u^\infty\leq 0.8$
and $0.1\leq St\leq 100$,
where we observe fixed, bubbling and channeling phases.
Transitions among these phases will be discussed below.

In the fixed phase at low flow rate,
particles are rest at the bottom.
At the critical flow rate $u^\infty=u_c$,
the particles begin to be fluidized.
It seems that the transition between the fixed phase and the fluidized phase
is independent of $St$.
We observe two fluidized phases.
One is {\it the channeling phase} observed for small $St$
where we can see a channel or a path of fluid flow.
Another is {\it the bubbling phase} observed for large $St$
where bubbles raise through the particle beds.
Typical snapshots of them
are shown in Figs. \ref{fig:config-channel} and \ref{fig:config-bubble}.
We show the area of channel-bubble transition observed in our simulations
as the transition area in Fig. \ref{fig:parameter-space},
where we will see that
statistical quantities qualitatively change their behaviors.

To characterize the transition of fluidization quantitatively,
we calculate the kinetic energy per particle
$E(t)$ defined by
\begin{equation}
  E(t)
  =
  {1\over N_{\text{M}}}\sum_{\alpha=1}^{N_{\text{M}}}
  |{\bf U}^\alpha(t)|^2.
\end{equation}
A typical behavior of $E(t)$ in the bubbling phase is shown
in Fig. \ref{fig:eng-time}.
We observe regular behavior
after the minimum point following the first peak of $E(t)$.
In the channeling phase,
we also observe qualitatively similar behavior of $E(t)$
to Fig. \ref{fig:eng-time},
though the period of peaks is smaller and each peak is not distinguishable.
We now introduce the average of $E(t)$
defined by
\begin{equation}
  \bar{E}
  =
  {1\over\Delta{\cal T}}\int_{\Delta{\cal T}}dt\ E(t),
\end{equation}
where $\Delta{\cal T}$ is the period of regular behavior in $E(t)$.
From Fig. \ref{fig:eng-u},
we observe a continuous transition at $u^\infty=u_c$
and the linear behavior of $\bar{E}(u^\infty)$
in the fluidized phase ($u^\infty>u_c$).
This behavior can be well fitted by
\begin{equation}
  \label{eq:eng-u-fit}
  \bar{E}(u^\infty)
  =
  \left\{
    \begin{array}{cr}
      0 & (u^\infty<u_c)\\
      A_E(u^\infty-u_c) & (u^\infty>u_c)
    \end{array}\right.,
\end{equation}
where $A_E$ and $u_c$ are the fitting parameters
which depend on $St$.
Equation (\ref{eq:eng-u-fit}) defines $u_c$ shown in
Fig. \ref{fig:parameter-space}.

Now we show that
the transition of fluidization can be understood
as the process generating the free volume
around the particles in the fixed phase.
It is useful to remember that
our model is 
Galilei invariant,
that is,
the system
with the fixed particles of ${\bf U_{\text{F}}}=\vec{0}$
under the flow rate $u^\infty$
is equivalent to
that
with the fixed particles of ${\bf U_{\text{F}}}=-u^\infty{\bf E}_z$
under the flow rate $0$.
Let us consider the process under the latter situation.
First we define $U_{\text{fall}}$
which is the falling velocity of the mobile particles in the fixed phase
without the support of fixed particles.
If the flow rate $u^\infty$ is smaller than $U_{\text{fall}}$,
the mobile particles cannot pass over the fixed particles
moving downward with $-u^\infty$.
Therefore the mobile particles hold on the fixed particles
and the gap between the mobile particles and  the fixed particles 
is not generated.
While the flow rate $u^\infty$ is larger than $U_{\text{fall}}$,
the mobile particles apart from the fixed particles
and then the gap between them is generated.
The gap causes the Rayleigh-Taylor instability observed when
a heavy fluid exists above a light fluid\cite{chandrasekhar}.
Then the gap may grow into a bubble
and propagate upward through the particles,
or may construct a channel.
From this discussion,
the critical flow rate $u_c$ is determined
by the falling velocity $U_{\text{fall}}$.
This suggests that $u_c$ is independent of $St$
because the falling velocity $U_{\text{fall}}$ can be evaluated
as the sedimentation rate of suspensions\cite{hayakawa1995}.

Next we discuss the channel-bubble transition.
In view of Figs. \ref{fig:config-channel} and \ref{fig:config-bubble},
it is hard to distinguish the channeling phase from the bubbling phase.
At first,
we show the variance $V_H$ defined by
\begin{equation}
  V_H
  =
  {1\over \Delta{\cal T}}
  \int_{\Delta{\cal T}}dt\ 
  \biglb( H(t)-\bar{H} \bigrb)^2,
\end{equation}
where $H$ and $\bar{H}$ are
the height of the center of mass and its average defined by
\begin{equation}
  H(t)
  =
  {1\over N_{\text{M}}}\sum_{\alpha=1}^{N_{\text{M}}}
  z^\alpha(t)
\end{equation}
and
\begin{equation}
  \label{eq:avg-height}
  \bar{H}
  =
  {1\over \Delta{\cal T}}
  \int_{\Delta{\cal T}}dt\ H(t).
\end{equation}
We expect that
$V_H$ is small in the channeling phase
and large in the bubbling phase.
Figure \ref{fig:cofm-vari-st} shows the corresponding behavior
for $u^\infty=0.3$
and the transition is observed around $St=5$.
For other cases,
the channel-bubble transitions are observed in the area
shown in Fig. \ref{fig:parameter-space}.

We also observe some qualitative changes in the physical quantities
corresponding to the channel-bubble transitions.
Here we discuss the behavior of $\bar{E}(St)$.
From Fig. \ref{fig:eng-st},
we see that
$\bar{E}$ increases with $St$ in the channeling phase,
and $\bar{E}$ decreases with $St$ in the bubbling phase.
The behavior 
in the channeling phase can be understood by the following reason.
At $St=0$,
we observe a steady channel, no relative motions among particles.
When $St$ increases,
the particles on the channel can be fluidized.
This is because $St$ is the relaxation time of
the particle velocity ${\bf U}$ to their terminal velocity ${\bf V}$,
which is determined under the case of $St=0$,
and the non-relaxed particles cause the collapse of the channel.
Therefore, $\bar{E}(St)$ increases with $St$ in the channeling phase.
On the other hand,
we observe that $\bar{E}$ decreases with $St$ in the bubbling phase.
This can be understood as follows.
From the previous paper \cite{ichiki1995} or in Fig. \ref{fig:eng-time},
we have seen that the relative motion of particles or the kinetic energy
is generated by bubbles
which have the difference of the local volume fraction in the system.
In fact, bubbles are always accompanied by the convective motion
of particles.
Because $St$ is the response time
of the velocity for the change of configurations,
bubbles become obscure as $St$ increases.
In fact, we see a sharp bubble and
the definite convective motion of particles
for $St=10$ (Fig. \ref{fig:cnf-b266f3-convex}),
while we see a relatively obscure bubble and the weaker convection
for $St=100$ (Fig. \ref{fig:cnf-b266g3-bubble}).
Therefore $\bar{E}(St)$ in the bubbling phase decreases with $St$.
As a result,
$\bar{E}(St)$ has a peak around the transition point.
This reflects on the $St$ dependence of various properties
such as $\mu_e(St)$ shown in Sec. \ref{sec:visc-einstein}
and $\bar{H}(St)$ discussed in Sec. \ref{sec:liquid}.

\subsection{The analogy to equilibrium systems}
\label{sec:visc-einstein}
In this section,
we demonstrate the existence of surprising correspondences
in statistical quantities
between fluidized beds which is in highly nonequilibrium states
and quasi equilibrium systems.
The result of $\bar{E}(u^\infty)$ in Fig. \ref{fig:eng-u}
suggests that
the flow rate $u^\infty$ behaves as the effective temperature
of the environment such as that of the heat bath for equilibrium systems.
Therefore, the critical flow rate $u_c$ may correspond
to the critical temperature
and $\bar{E}$ may be the order parameter of the fluidization.
In the following,
we will interpret the results of our simulations
using this effective temperature $u^\infty$.

In nearly equilibrium systems at temperature $T$,
the resistance $\zeta$ of a tracer particle is given by
\begin{equation}
  \zeta
  =
  {kT\over D},
  \label{eq:einstein}
\end{equation}
where $D$ is the diffusion constant and
$k$ is the Boltzmann constant\cite{einstein1905}.
This is the Einstein relation
and is the simplest form of the fluctuation-dissipation theorem
which relates the correlation functions in the equilibrium state
and the transport coefficients.

Similarly,
we introduce for our simulations the effective viscosity $\mu_e$ as
\begin{equation}
  \mu_e
  =
  {u^\infty\over D_p},
  \label{eq:einstein-fb}
\end{equation}
where $D_p$ is the diffusion constant of our simulations
defined by
\begin{equation}
  D_p
  =
  {1\over 2}{1\over N_{\text{M}}}\sum_{\alpha=1}^{N_{\text{M}}}
  |\tilde{\bf U}^\alpha(\omega=0)|^2,
\end{equation}
where $\tilde{\bf U}^\alpha(\omega)$ is the Fourier transform
of ${\bf U}^\alpha(t)$
calculated by the standard FFT.
Equation (\ref{eq:einstein-fb}) can be regarded
as an extension of Eq. (\ref{eq:einstein})
with the aid of the effective temperature $u^\infty$,
because the viscosity is usually proportional to the resistance.
The observed viscosity $\mu_e$ defined by (\ref{eq:einstein})
is shown in Figs. \ref{fig:visc-diff-u} and \ref{fig:visc-diff-st}.

If the definition of the viscosity (\ref{eq:einstein-fb}) is self-consistent,
the Einstein relation or the fluctuation-dissipation relation
in fluidized beds
may be valid with the replacement of $kT$ by $u^\infty$.
This statement is interesting
because the system is in a highly nonequilibrium state
and there is no reason of the existence of the fluctuation-dissipation relation
in the sense of linear nonequilibrium statistical mechanics.

We find that
the flow-rate dependence of the viscosity $\mu_e(u^\infty)$
obtained by Eq. (\ref{eq:einstein-fb}) obeys
the Arrhenius function
\begin{equation}
  \label{eq:arrhenius}
  \mu_e(u^\infty)
  \propto
  \ e^{\varepsilon/u^\infty},
\end{equation}
where $\varepsilon$ is a fitting parameter.
The fitting by Eq. (\ref{eq:arrhenius})
with $\varepsilon=0.113 \pm 0.017$ is also shown
in Fig. \ref{fig:visc-diff-u}.
We compare our result of $\mu_e$ with experimental result
in fluidized beds\cite{furukawa1958}.
In the experiment of fluidized beds,
the shear viscosity measured by the modified Stormer viscometer
also obeys the Arrhenius function of Eq.(\ref{eq:arrhenius})
\cite{furukawa1958}.
Therefore,
our result from Eq. (\ref{eq:einstein-fb}) is consistent
with the experiment.
This behavior which can be understood by a dense liquid theory in part
will be explained in Sec. \ref{sec:liquid}.

The connection between the non-Gaussian property and
the dissipation in the system
have been discussed for granular materials
\cite{taguchi1995b,brey1996,ichiki1995}.
Here we check the non-Gaussian property in
the velocity distribution functions $P(U_x)$ obtained from our simulations.
To characterize the non-Gaussian property of the velocity distribution,
we calculate the 4th cumulant $C_4$ defined by
\begin{eqnarray}
  C_4(U_x)
  &=&
  \langle {U_x}^4\rangle
  -3\langle {U_x}^2\rangle ^2
  -4\langle {U_x}\rangle\langle {U_x}^3\rangle
  \nonumber\\
  &&
  +12\langle {U_x}\rangle^2\langle {U_x}^2\rangle
  -6\langle {U_x}\rangle ^4.
\end{eqnarray}
The resultant behaviors of $C_4$ are shown in
Figs. \ref{fig:cum-u} and \ref{fig:cum-st},
where they are scaled by the square of variance (or 2nd cumulant)
defined by
\begin{equation}
  C_2({U_x})
  =
  \langle {U_x}^2\rangle -\langle {U_x}\rangle ^2.
\end{equation}
This non-Gaussian parameter $C_4/(C_2)^2$
is zero for the Gaussian distribution
and $3$ for the exponential distribution.
These behaviors of $C_4/(C_2)^2$
in Figs. \ref{fig:cum-u} and \ref{fig:cum-st}
are similar to those of the effective viscosity $\mu_e$
in Figs. \ref{fig:visc-diff-u} and \ref{fig:visc-diff-st}.
In fact,
we can also fit $C_4/(C_2)^2$ by the Arrhenius function
\begin{equation}
  \label{eq:cum-u-fit}
  C_4/(C_2)^2
  \propto
  \exp(\varepsilon'/u^\infty),
\end{equation}
as shown in Fig. \ref{fig:cum-u} with $\varepsilon' =0.175\pm 0.045$.
It will be an interesting subject
that we will study a quantitative relation between
(\ref{eq:arrhenius}) and (\ref{eq:cum-u-fit}).

\section{Discussions}
\label{sec:liquid}
In this section,
we demonstrate that the qualitative understanding of the results
of our simulation is possible
with the aid of the hole theory applied to simple liquids.

First we discuss the flow rate dependence of the viscosity $\mu_e(u^\infty)$.
The hole theory, which is used for the behavior of simple liquids,
is based on the following picture.
A molecule in a liquid can move when a free volume or a hole
is generated around it.
From this picture,
the empirical relation of the viscosity $\mu_l$ \cite{frenkel} is derived as
\begin{equation}
  \label{eq:andrade}
  \mu_l(T)
  \propto
  \exp\left({\varepsilon_l\over kT}\right),
\end{equation}
where $\varepsilon_l$ is the activation energy to create the free volume.
If we regard $u^\infty$ as the temperature,
Eq. (\ref{eq:arrhenius}) is identical to Eq. (\ref{eq:andrade}).
This suggests that the averaged behavior in
the fluidized beds may be understood
by the hole theory in simple liquids
\cite{hayakawa1997,warr-pre,ristow-pre}.

Next we discuss $St$ dependence of the
height of center of mass $\bar{H}(St)$
defined by (\ref{eq:avg-height}).
In Fig. \ref{fig:cofm-st}
we show a typical behavior of $\bar{H}(St)$ at $u^\infty=0.3$.
We see the qualitative change of behavior in the transition area
in Fig. \ref{fig:parameter-space},
where $\bar{H}$ is almost constant in the channeling phase,
and $\bar{H}$ increases logarithmically in the bubbling phase.
Since the change of behavior with $St$ in the channeling phase
is only how the channel collapses,
$\bar{H}(St)$ is expected to be independent of $St$.
While the behavior in the bubbling phase
is interesting.
We can fit the data in Fig. \ref{fig:cofm-st} as
\begin{equation}
  \label{eq:cofm-st-fit}
  \bar{H}(St)
  =
  C_H\log(St)+D_H,
\end{equation}
where the fitting parameters $C_H$ and $D_H$ depend on $u^\infty$.
Equation (\ref{eq:cofm-st-fit}) can be rewritten as
\begin{equation}
  \label{eq:st-exp-cofm}
  St
  =
  \exp\left({\bar{H}-D_H\over C_H}\right).
\end{equation}
Here $\bar{H}-D_H$ can be understood as the volume expansion $\Delta V$,
since $D_H$ is the height at $St=1$.
We show $u^\infty$ dependence of $C_H$ in Fig. \ref{fig:cofm-st}.
From this figure,
$C_H$ is approximately proportional to the effective temperature $u^\infty$.
Since $St$ is the characteristic time $\tau$ of the system,
we rewrite Eq.(\ref{eq:st-exp-cofm}) as
\begin{equation}
  \label{eq:time-free-vol}
  \tau
  \propto
  \exp\left(F{\Delta V\over u^\infty}\right),
\end{equation}
where $F$ is a constant.
Equation (\ref{eq:time-free-vol}) is consistent with Eq.(\ref{eq:andrade})
under the reasonable assumption where
the viscosity is characterized by the time $\tau$
to generate the free volume,
and the activation energy to generate the free volume or expansion
is proportional to $\Delta V$.

Before closing this section, we give some remarks.
Although the transition of fluidization in experiments seems to be
the discontinuous phase transition\cite{discontinuous-fluidization},
our simulations suggests a continuous phase transition.
Also it is an open problem that
at present we cannot reproduce {\it homogeneous phase}
in our simulation.
For these problems,
we need to examine carefully the difference
between the experiments and the simulations,
and we must investigate the behavior near the critical flow rate $u_c$
in detail,
because the discontinuous phase transition and the homogeneous phase
are observed there in experiments.

\section{Conclusions}
In this paper,
we have carried out systematic simulations
with the change of two control parameters,
the flow rate of the fluid $u^\infty$ and
the effective Stokes number $St$.
When the flow rate $u^\infty$ is small,
particles rest in {\it the fixed phase}.
Above the critical flow rate $u_c$, particles are fluidized.
The critical value $u_c$ is independent of $St$.
We have found two fluidized phases, {\it the channeling phase}
and {\it the bubbling phase},
where the former changes to the latter as $St$ increases.

We have found that
the flow rate $u^\infty$ plays the role of the effective temperature.
In terms of the effective temperature $u^\infty$,
we have defined the effective viscosity $\mu_e$
with the aid of the Einstein relation.
The flow-rate dependence of the viscosity $\mu_e$ is
similar to that in the experiments in real fluidized beds.
We also find that
the viscosity $\mu_e(u^\infty,St)$ can be an index of
the non-Gaussian property in velocity distribution of particles.
This property is consistent with the behavior on granular materials
or the system of inelastic particles.
Qualitative behavior of fluidized beds
such as $\mu_e(u^\infty)$ and $\bar{H}(St)$ can be understood
by means of the hole theory which has been used for simple liquids.

\acknowledgments
We would like to thank S. Sasa, Y-h. Taguchi, T. Ooshida
for fruitful discussions on the subject of this work.
K.I. also thanks M. Doi and J. F. Brady for encouragements and
useful comments on this work.
This work has been partially supported by JSPS Research Fellowships for
Young Scientists
and Grant in Aid of Ministry of Education of
Science and Culture in Japan.
Numerical simulations was carried out by the facilities of
the Supercomputer Center at the Institute for Solid State of Physics
at the University of Tokyo.

\appendix
\section{The ideal model of fluidized beds}
Here
we review our model of fluidized beds,
presented in the previous paper\cite{ichiki1995},
where only essences are extracted from real systems
and all other irrelevant mechanisms are neglected.

\subsection{Equation of motion}
\label{sec:eofm}
We construct the model by only four mechanisms,
which are the inertial effect of the particles,
hydrodynamic interaction through the fluid,
the gravitational force and the contact force.
Therefore the equation of motion can be written as
\begin{equation}
  \label{eq:eofm}
  St_0{d {\bf U}\over dt}
  =
  {\bf F}_f
  +{\bf F}_g
  +{\bf F}_c,
\end{equation}
where ${\bf F}_f,{\bf F}_g,{\bf F}_c$
are the force from the fluid, the gravitational force and the contact force
respectively.
$St_0$ is the bare Stokes number defined by
\begin{equation}
  St_0
  =
  {mU_0\over 6\pi\mu a^2}.
\end{equation}
Particles are assumed to be monodisperse and hard-core spheres
and rotational motions and torques acting on particles are neglected.

The gravitational force ${\bf F}_g$ can be written as
\begin{equation}
  {\bf F}_g
  =
  -{\bf E_z},
\end{equation}
where ${\bf E}_z$ is the unit vector directed to the $z$ axis
by the notation of Eq. (\ref{eq:n-vector}).
It is assumed that the direction of the gravity is $-z$.

It is assumed that ${\bf F}_c$ is
the impulse by prefect elastic collisions.
Because collisions are inelastic in real systems,
this assumption of elastic collision
means our standpoint of modeling
that the essential mechanism of fluidized beds
is not the inelasticity in collisions but the hydrodynamic interaction.
Even in this case,
the model is dissipative,
because the hydrodynamics interaction is nothing less than the friction.
In our simulation,
we represent the collision by the momentum exchange at contact
in stead of the contact force ${\bf F}_c$.
Therefore we do not write ${\bf F}_c$ explicitly in the equations
in the following discussion.

\subsection{Hydrodynamic interaction}
\label{sec:hydrodynamic}
In our model,
hydrodynamic interaction among particles through the fluid
is considered under the Stokes approximation
where the viscous effect of the fluid dominates the inertia of the fluid.
The reason to adopt the Stokes approximation is as follows:
The hydrodynamic interaction is
the friction between the particles and the fluid
and the friction is originated by the viscosity of the fluid.

In the Stokes approximation,
the force acting on the particles from the fluid ${\bf F}_f$
and the velocity of the particles ${\bf U}$ are related
by the resistance matrix $\tensor{R}$ as
\begin{equation}
  \label{eq:resistance}
  -{\bf F}_f
  =
  \tensor{R}
  \cdot
  ({\bf U}-{\bf u}^\infty),
\end{equation}
where ${\bf u}^\infty$ is the velocity of the fluid without particles.
$\tensor{R}$ contains all information about the interaction
and depend only on the configuration of particles.

To calculate $\tensor{R}$,
we adopt the method in the Stokesian dynamics,
which is developed by J.F.Brady and his collaborators
for dense colloidal particles
\cite{durlofsky1987a,brady1988b,brady1988a}.
In the Stokesian dynamics,
$\tensor{R}$ is constructed
from the two contributions in limited cases,
which are the mobility matrix in dilute limit ${\tensor{M}}^\infty$
and the exact resistance matrix in two-body problem ${\tensor{R}}_{\text{2B}}$
\cite{jeffrey1984}, as follows,
\begin{equation}
  \tensor{R}
  =
  ({\tensor{M}}^\infty)^{-1}
  +{\tensor{R}}^{\text{lub}},
  \label{eq:stokesian}
\end{equation}
where ${\tensor{R}}^{\text{lub}}$ is constructed by the pairwise-additive manner
from the two-body lubrication matrix ${\tensor{R}}^{\text{lub}}_{\text{2B}}$
which is defined by
\begin{equation}
  {\tensor{R}}^{\text{lub}}_{\text{2B}}
  =
  {\tensor{R}}_{\text{2B}}
  -({\tensor{M}}^\infty_{\text{2B}})^{-1}.
\end{equation}
${\tensor{M}}^\infty_{\text{2B}}$ is the two-body mobility matrix
in dilute limit.
In general ${\tensor{M}}^\infty$ is formulated by the multipole expansion
\cite{durlofsky1987a}.
In the model of fluidized bed, however,
we need to introduce the effect of the chamber.
The chamber in the real fluidized beds has two contributions
which are to bound the fluid by the vertical wall
and to support the particles by the bottom.

We introduce the contribution of bounding the fluid
in terms of the periodic boundary condition.
Because ${\tensor{M}}^\infty$ has the long-range interaction,
we use the Ewald summation technique\cite{beenakker1986}.
We can construct the resistance matrix under the periodic boundary condition
also by Eq. (\ref{eq:stokesian})
only replacing ${\tensor{M}}^\infty$ to the Ewald summed tensor
for $N$ particles in the unit cell\cite{brady1988b}.

On the other hand,
the contribution of the bottom supporting the particles
is introduced by the particles fixed in space.
In this case, we get the force from the fluid to the mobile particles
${\bf F}_f$ in Eq. (\ref{eq:eofm}) as follows,
\begin{equation}
  -{\bf F}_f
  =
  \tensor{R}_{\text{MM}}
  \cdot
  ({\bf U}_{\text{M}}-{\bf u}^\infty)
  -
  \tensor{R}_{\text{MF}}
  \cdot
  {\bf u}^\infty,
  \label{eq:force-fluid-fixed}
\end{equation}
where
the subscripts ``M'' and ``F'' represent
{\it mobile} and {\it fixed} particles respectively.
The complete form of the resistance relation is
\begin{equation}
  -\left[
  \begin{array}{c}
    {\bf F}_f\\
    {\bf F}_{\text{F}}
  \end{array}\right]
  =
  \left[
  \begin{array}{cc}
    \tensor{R}_{\text{MM}} &
    \tensor{R}_{\text{MF}} \\
    \tensor{R}_{\text{FM}} &
    \tensor{R}_{\text{FF}}
  \end{array}\right]
  \cdot
  \left[
  \begin{array}{c}
    {\bf U}_{\text{M}}-{\bf u}^\infty\\
    \vec{0}-{\bf u}^\infty
  \end{array}\right],
\end{equation}
where ${\bf F}_{\text{F}}$ is the force acting on the {\it fixed} particles.

For simplicity
we only discuss the case without fixed particles
in the following.
However we can get the correct forms
only the replacement of ${\bf F}_f$ of (\ref{eq:resistance})
by (\ref{eq:force-fluid-fixed}).

\subsection{Effective inertia}
The inertia of particles causes the relaxation process on the velocity
from the initial value to the optimal value
where
the inertia corresponds to the relaxation time.
The optimal value is usually called {\it the terminal velocity}
determined on the steady state.
For the fluidized beds,
the terminal velocity ${\bf V}$ is determined
by Eq. (\ref{eq:eofm}) with $d{\bf U}/dt=\vec{0}$
and get Eq.(\ref{eq:term-vel}).
This is the case that all forces acting on particles balance.
With this terminal velocity
we can write the relaxation process as
Eq.(\ref{eq:eofm-efct}).

Equation (\ref{eq:eofm-efct}) is the same as used in the previous paper
\cite{ichiki1995}.
Although we had argued that this equation might be justified
in some approximation,
we would state here that (\ref{eq:eofm-efct})
contains all of essential processes in fluidized beds.
On this point, we will discuss in Appendix \ref{sec:inertia}.

\section{The inertial effect of particles}
\label{sec:inertia}
Here we discuss the difficulties
arising when we use Eq. (\ref{eq:eofm}) with ${\bf F}_f$
in the Stokes approximation (\ref{eq:resistance}).

If we write Eq. (\ref{eq:eofm}) with ${\bf F}_f$, we get
\begin{equation}
  \label{eq:eofm-direct}
  St_0
  {\tensor{R}}^{-1}\cdot
  {d\over dt}{\bf U}
  =
  -{\bf U}+{\bf V}.
\end{equation}
From the simulation of Eq.(\ref{eq:eofm-direct}),
we observe no collision between particles even in the case with large $St_0$.
Particles form a cluster and
relative motions among them almost disappear.

This situation may be understood by the following model,
\begin{equation}
  \label{eq:eofm-direct-simple}
  St_0{d U\over dt}
  =
  -{1\over r-2a}U.
\end{equation}
Here we extract the radial component of the motion between
two particles.
$r$ denotes the separation between the centers of the pair
and $U$ denotes the relative velocity.
The resistance $1/(r-2a)$ reflects the lubrication effect
which diverge at contact ($r\rightarrow 2a$).
In this model, we need  initially the infinite energy
to approach the contact point ($r=2a$)
even in the case of large $St_0$.
Thus the model (\ref{eq:eofm-direct-simple}) cannot contains any collision.

Our model (\ref{eq:eofm-efct}) can be understood
as the renormalization of the singularity in the lubrication,
because we get Eq. (\ref{eq:eofm-efct})
by multiplying the singularity $\tensor{R}$
on the inertial term of (\ref{eq:eofm-direct}).
(Here we note that $\tensor{R}$ is dimensionless
because it scaled by $6\pi\mu a$.)
Our model behave reasonably
like the real fluidized beds
where collisions are occurred so frequently.
This suggests that the singularity of the lubrication
must be prevented from some mechanisms in the real systems.

We can imagine several possible mechanisms preventing the singularity.
For example, if there are some dimples on the surface of the particles,
they collide before the mean surfaces contact.
From another point of view we can also say that 
the continuous description of the fluid in the gap between the particles
breakdowns
when particles approach closely and the gap becomes comparable to
the mean-free path of the molecules of the fluid\cite{sundararajakumar1996}.

Recently a model in this context has been presented\cite{sangani1996}.
They introduce a cut-off length,
which may correspond to the height of the dimples on the surface
or the mean-free path of the fluid molecules.
If particles approach with each other within the cut-off length,
the gap between them is assumed to be the cut-off
in calculation of hydrodynamic interaction.
The result of their simulation is suggestive
even though the situation,
which is the behavior in the shearing flow without the gravity,
is different.
Their results are characterized by the parameter
$St_s/\langle R_c\rangle$, where
$St_s=m\gamma/6\pi\mu a$ is the Stokes number in shear flow
with the shear rate $\gamma$
and $\langle R_c\rangle$ is the averaged resistance of a tracer
with the same volume fraction and the cut-off length.

From the above discussion,
we get the meaning of the effective Stokes number $St$ as
\begin{equation}
  St
  =
  St_0/\langle R_c\rangle ,
  \label{eq:st-mean}
\end{equation}
which depends on the cut-off length of the real systems.
We need more delicate investigations
for the dependence of the cut-off length
or the mechanism preventing the singularity in the lubrication.

\begin{figure}
  \caption{Simulations performed in the parameter space $(u^\infty,St)$
    with $N_{\text{M}}=256,N_{\text{F}}=10$ and $(L_x,L_y,L_z)=(34,2,100)$.
    In this figure we show the observed phases,
    {\it fixed phase} ($\bullet$),
    {\it channeling phase} ($\Box$),
    {\it bubbling phase} ($\circ$)
    and transition phase ($\triangle$).
    We also show the transition line of fluidization $u_c$ with solid line
    and the area of the channel-bubble transition with dotted line,
    which are discussed in the text.}
  \label{fig:parameter-space}
\end{figure}

\begin{figure}
  \caption{Typical snapshots of channeling phase
    with $St=0.5,u^\infty=0.15$.
    The time proceeds from left to right with the interval
    $20$ dimensionless time.}
  \label{fig:config-channel}
\end{figure}
\begin{figure}
  \caption{Typical snapshots of bubbling phase
    with $St=10.0,u^\infty=0.15$.
    The time proceeds from left to right with the interval
    $20$ dimensionless time.}
  \label{fig:config-bubble}
\end{figure}

\begin{figure}
  \caption{A typical behavior of $E(t)$ in the fluidized phase.
    Parameters used are $St=20, u^\infty=0.3$.
    The oscillation corresponds to the generation of bubbles.
    $1$ step means $1$ dimensionless time.}
  \label{fig:eng-time}
\end{figure}

\begin{figure}
  \caption{The flow rate dependence of the averaged energy
    $\bar{E}(u^\infty)$ with $St=10$.
    Error bars are their standard deviation.
    We can see that
    the transition of fluidization at $u^\infty=u_c$
    and the linear behavior of $\bar{E}$ for the flow rate $u^\infty$
    in the fluidized phase $(u^\infty>u_c)$.}
  \label{fig:eng-u}
\end{figure}

\begin{figure}
  \caption{$St$ dependence of
    the variance $V_H$ with $u^\infty=0.3$.
    We can see the transition around $St=5$.
    At the channeling phase
    the variance is small,
    while at the bubbling phase
    the variance become large.}
  \label{fig:cofm-vari-st}
\end{figure}

\begin{figure}
  \caption{$St$ dependence of the averaged energy $\bar{E}(St)$
    with $u^\infty=0.3$.
    We can observe a peak around the channel-bubble transition
    shown in Fig. \protect\ref{fig:parameter-space}.}
  \label{fig:eng-st}
\end{figure}

\begin{figure}
  \caption{The convective motion of particles in the bubble
    with $u^\infty=0.3,St=10$.
    Here two periodic images are shown.
    We can observe the sharp edge of the bubble
    and the definite convection.}
  \label{fig:cnf-b266f3-convex}
\end{figure}
\begin{figure}
  \caption{The convective motion of particles in the bubble
    with $u^\infty=0.3,St=100$.
    Here two periodic images are shown.
    Comparing to the Fig. \protect\ref{fig:cnf-b266f3-convex},
    we can observe broader edge of the bubble
    and weaker convection,
    where the scale of the velocities are the same.}
  \label{fig:cnf-b266g3-bubble}
\end{figure}

\begin{figure}
    \caption{Effective viscosity $\mu_e(u^\infty)$.
      This result is calculated on the simulation with $St=10.0$.
      The fitting by the Arrhenius type function (\protect\ref{eq:arrhenius})
      with $\varepsilon=0.113 \pm 0.017$ is also shown.}
    \label{fig:visc-diff-u}
\end{figure}
\begin{figure}
    \caption{Effective viscosity $\mu_e(St)$.
      This result is calculated on the simulation with $u^\infty=0.3$.}
    \label{fig:visc-diff-st}
\end{figure}

\begin{figure}
  \caption{The non-Gaussian parameter $C_4/(C_2)^2$
    with $St=10$ for the change of $u^\infty$.
    We also show the fitting by $\exp(\varepsilon'/u^\infty)$
    with $\varepsilon' =0.175\pm 0.045$.}
  \label{fig:cum-u}
\end{figure}
\begin{figure}
  \caption{The non-Gaussian parameter $C_4/(C_2)^2$
    with $u^\infty=0.3$ for the change of $St$.}
  \label{fig:cum-st}
\end{figure}

\begin{figure}
  \caption{$St$ dependence of $\bar{H}$
    with $u^\infty=0.3$.
    For the bubbling phase,
    we fit by Eq.(\protect{\ref{eq:cofm-st-fit}})
    with $C_H = 3.087 \pm 0.055,D_H = 16.51 \pm 0.14$.}
  \label{fig:cofm-st}
\end{figure}
\begin{figure}
  \caption{$u^\infty$ dependence of the fitting parameter $C_H$.
    We can see $C_H$ is linear for $u^\infty$.}
  \label{fig:cofm-fit-c}
\end{figure}

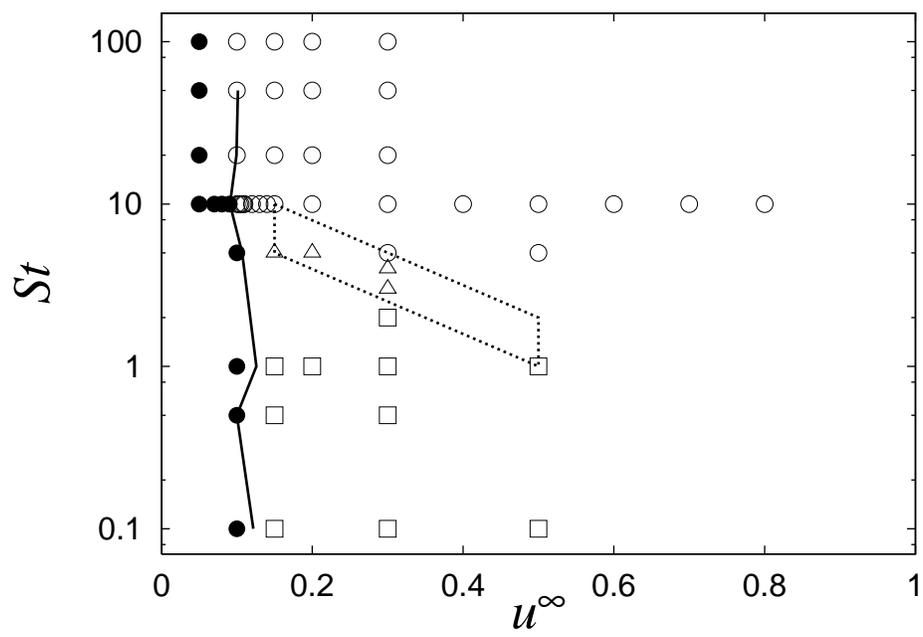
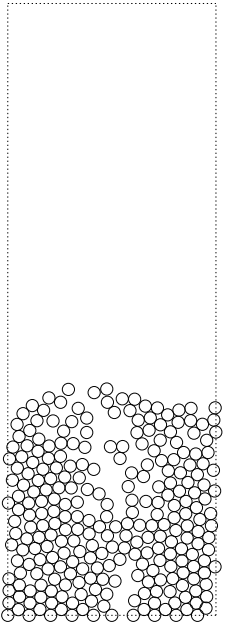
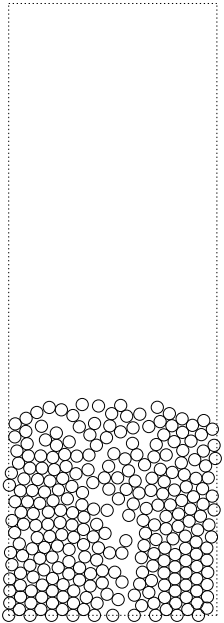
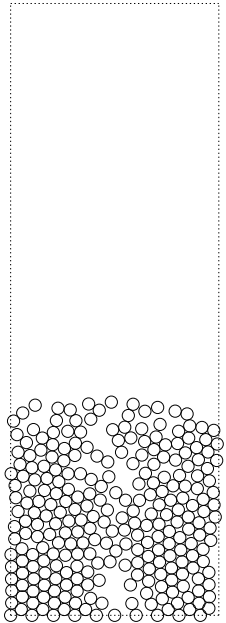
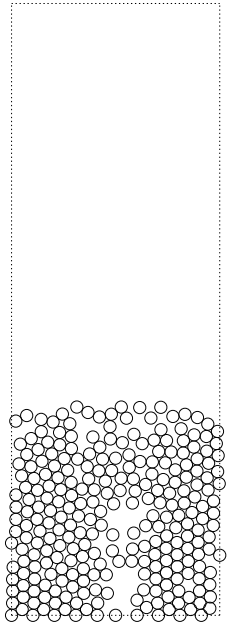
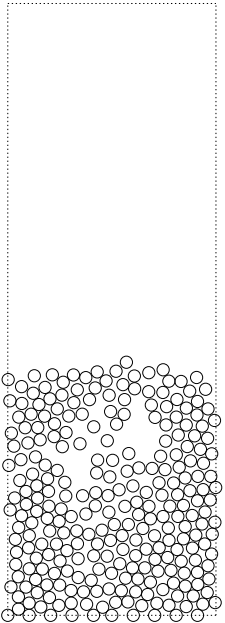
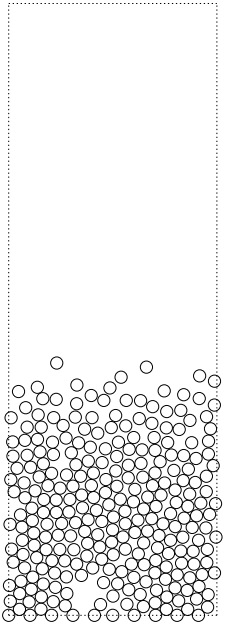
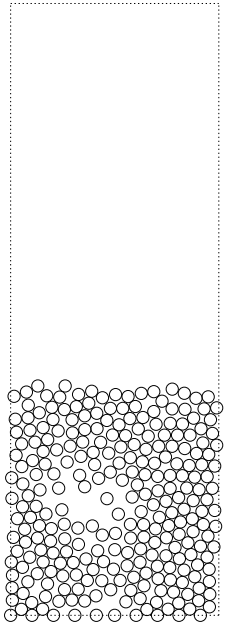
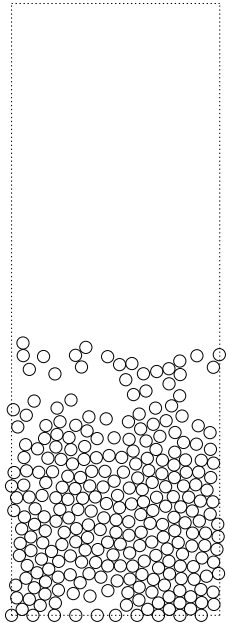
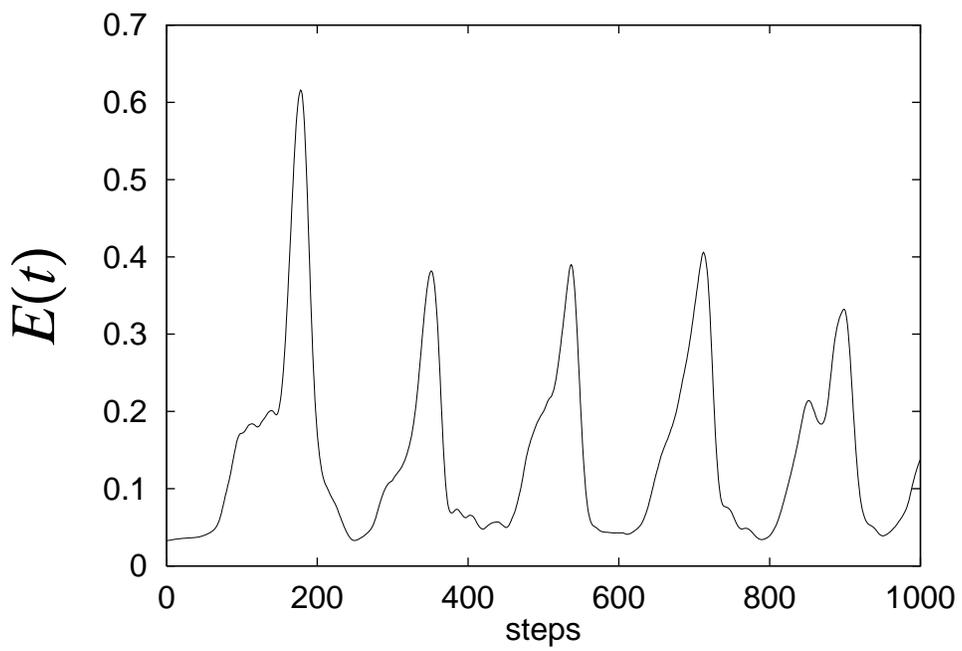
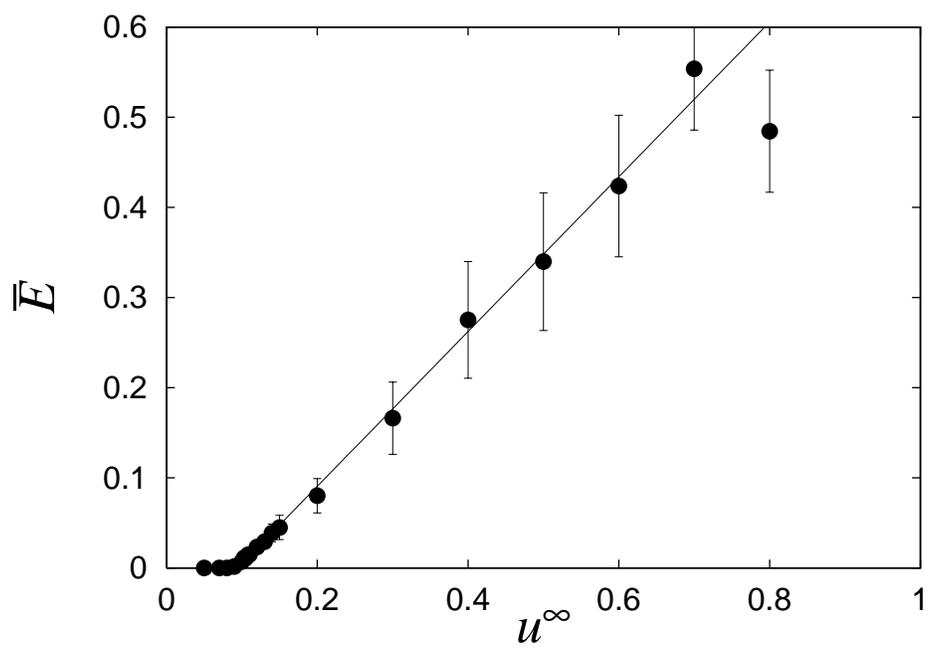
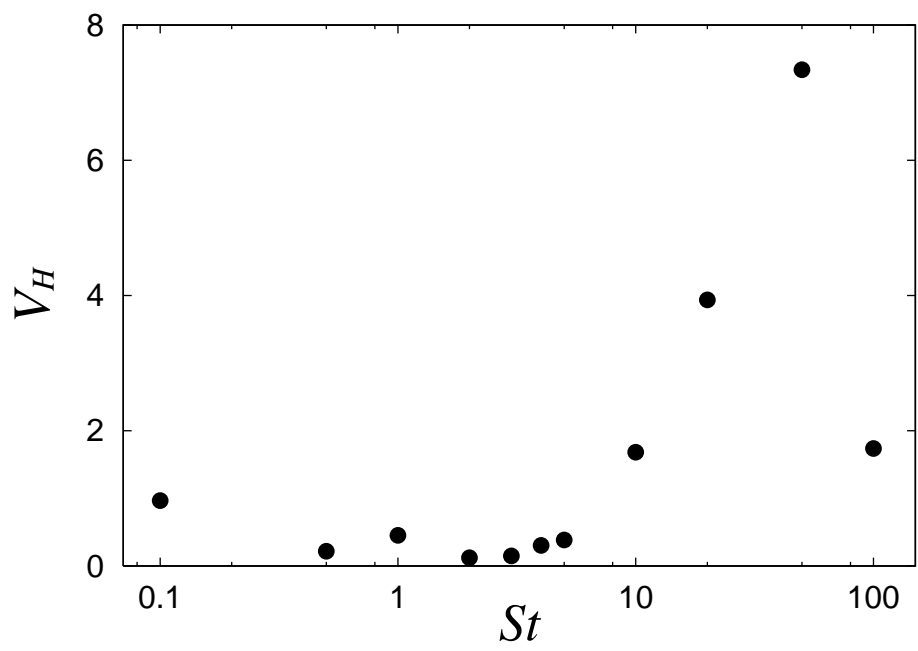
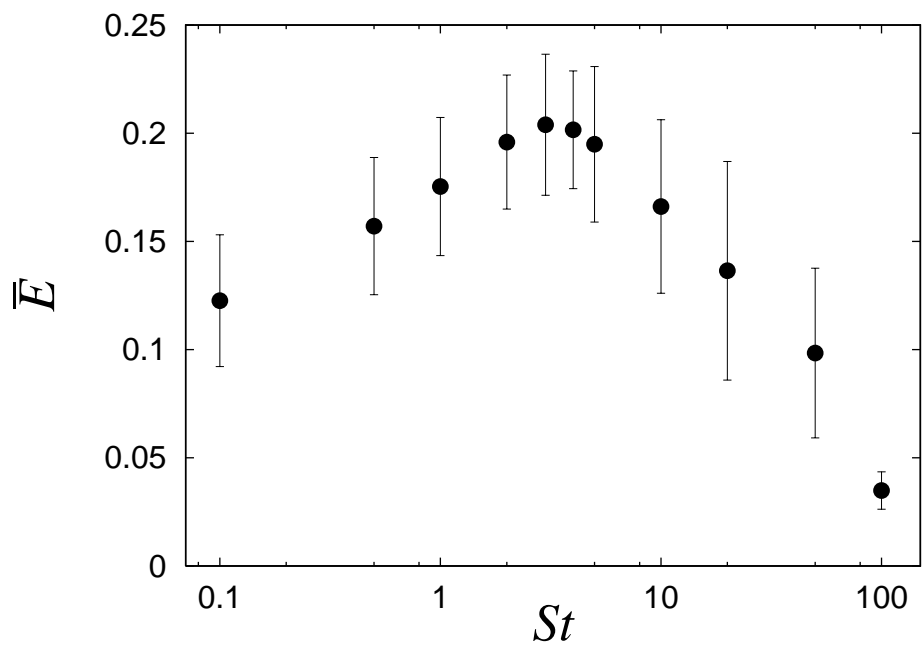
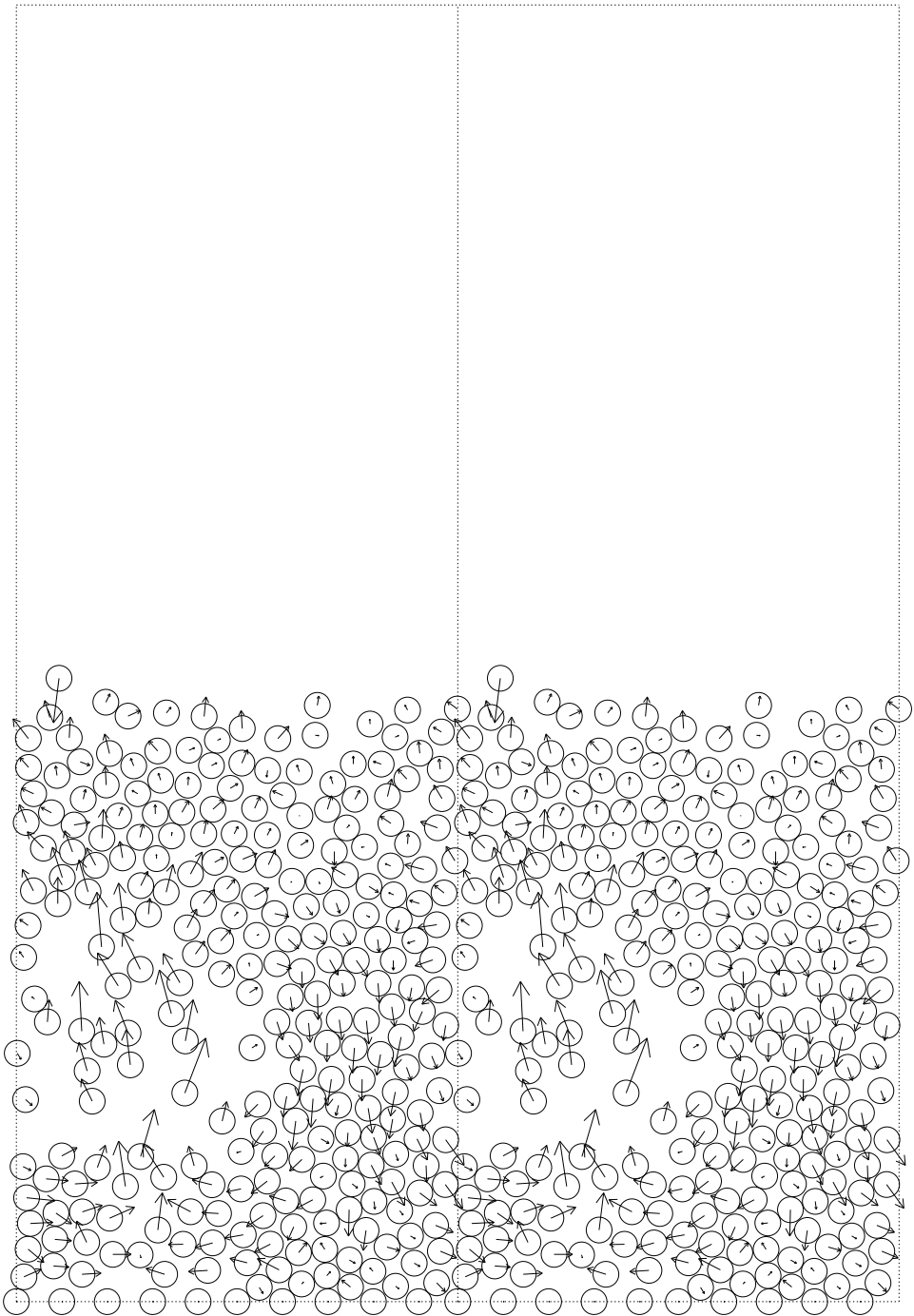
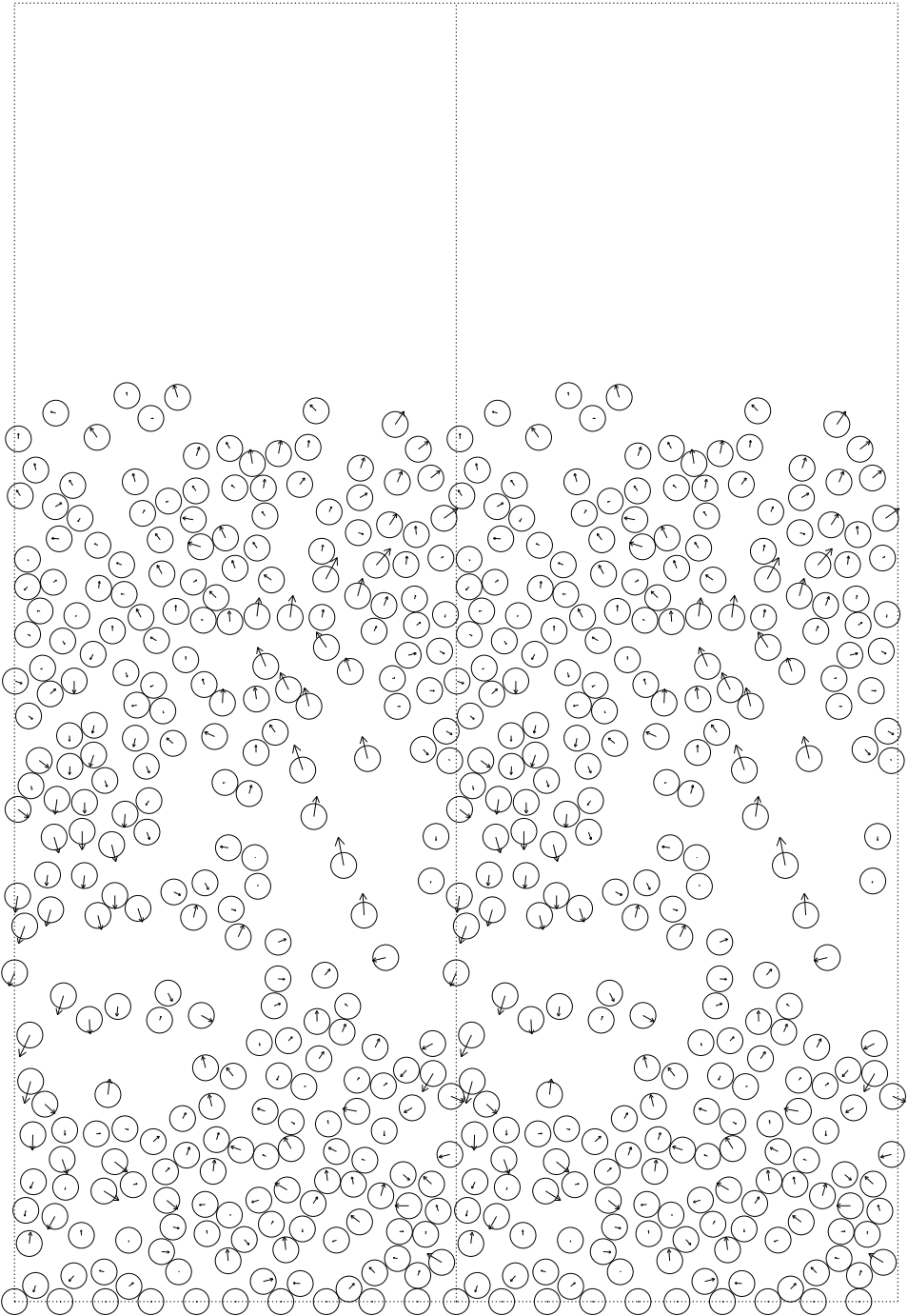
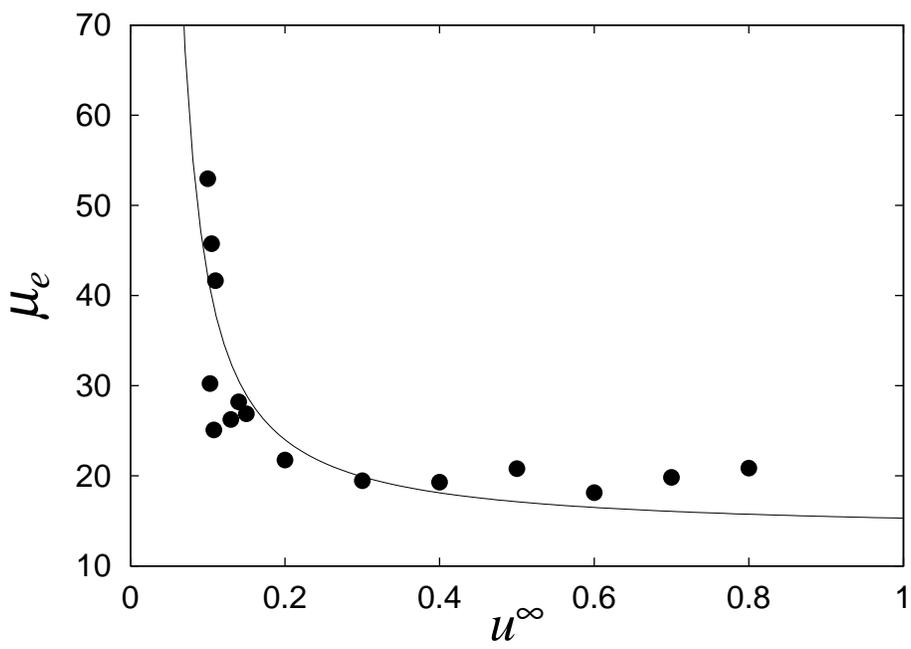
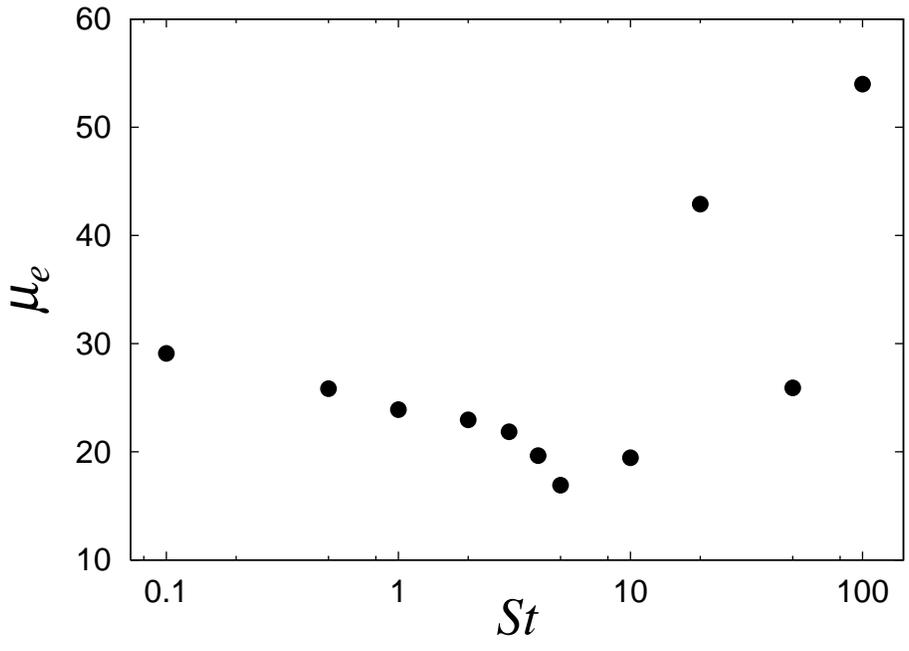
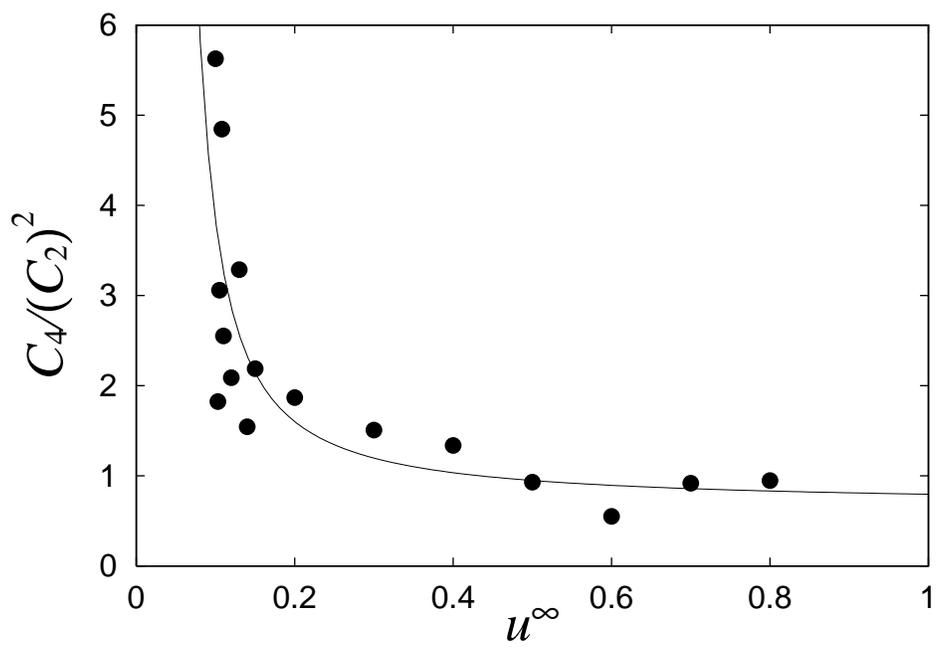
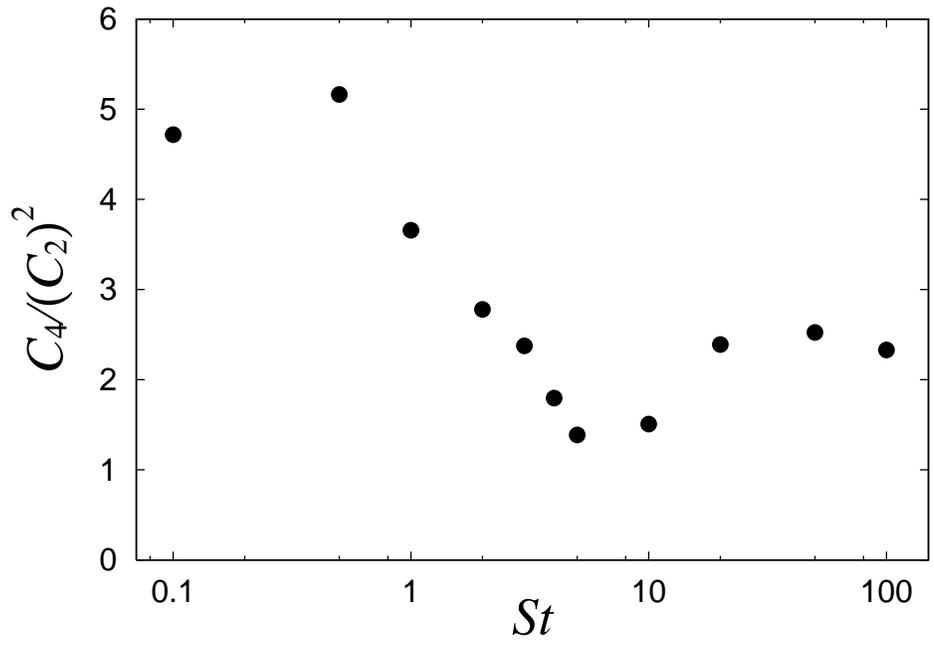
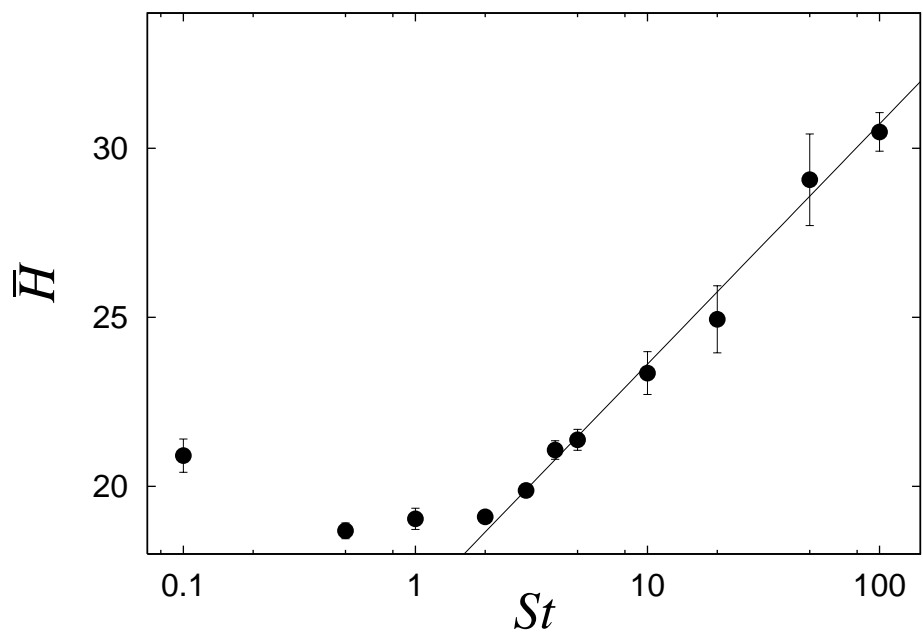
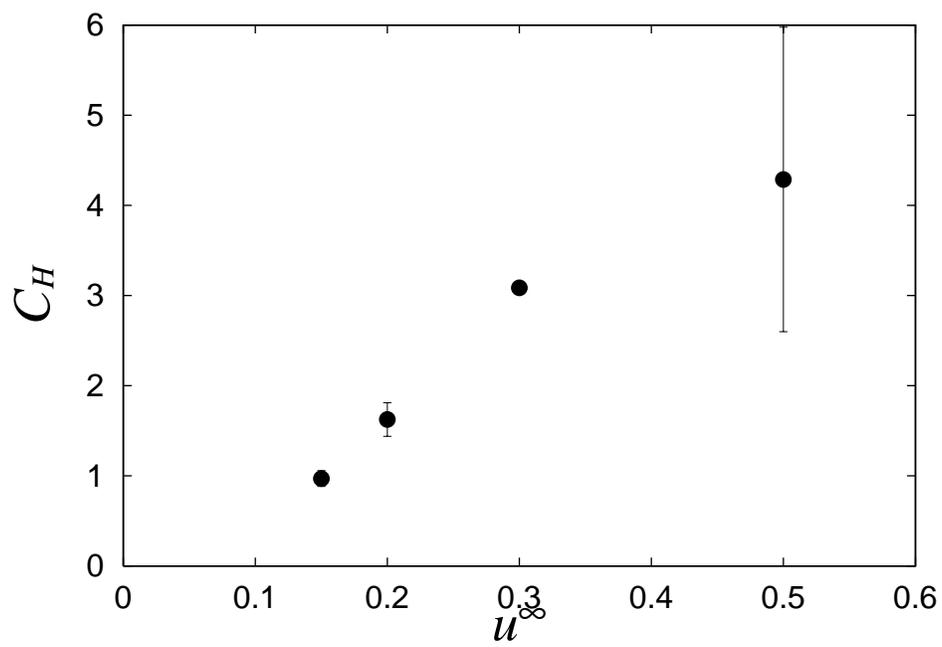
\end{document}